# The transition and coexistence of quantum droplets and solitons in quasi-1D dipolar Bose gas

Y. Y. Shi

*School of Physics and Electronic Information Engineering, Qinghai Normal University, Xining, Qinghai 810008, China*

In our study, we investigated bright solitons, dark solitons, and quantum droplets in quasi-one-dimensional dipolar Bose gases, and further validated the crossover and coexistence of quantum droplets and solitons using the Lieb-Liniger energy within the framework of local density approximation. Increasing the particle number transforms the bright dipolar soliton state into a stable self-bound quantum droplet state, with further increases in leading to a broader quantum droplet that enables the presence of dark solitons within it. This article is incomplete.

## I. Introduction

Quantum droplets, as an important component of research in ultracold atomic physics, have been originally observed in ultracold dipolar gas [1–4] and afterwards in Bose-Bose mixtures [5–7]. The collape-preventing mechanism is due to the quantum fluctuations, not accounted for in the original Gross-Pitaevskii equation (GPE), but corrected by a seminal Lee-Huang-Yang (LHY) term via introducing an additional repulsion into the ground state energy [8].The dimension-dependent nature of the LHY term results in significantly different equations for describing quantum droplets in different dimensions [9–12]. in addition to quantum droplets in 1D, 2D, and 3D, previous studies have investigated quantum droplets in cross dimensions [13–16] and in a ring geometry [17]. he extended GPE (EPGP) solely with the LHY correction does not yield a quantitative agreement with the quantum Monte Carlo predictions for strong interactions [8, 18]. By referencing the energy functional discussed by Lieb [19] and employing the local density approximation, reference [20] has bridged the gap in strong interactions within quantum droplets by introducing the Lieb-Liniger GPE (LLGPE), which has been demonstrated to effectively describe Bose-Einstein condensates (BEC) [14, 21–28]

On the one hand, bright solitons and quantum droplets exist in very different regimes, however, previous studies have shown that there is a transition between these two self-bound states [7, 16, 29–31]. On the other hand, the possibility of quantum droplet-dark soliton coexistence has been proven by recent research [15, 16].

In this work, we wish to explore the transition and coexistence of quantum droplets and solitons in the field of ultracold atoms. By increasing the particle number N, the bright dipolar soliton state gradually evolves into a stable self-bound quantum droplet state. At this point, further increasing the particle number N does not change the density of the quantum droplet but results in an extremely broad quantum droplet, providing conditions for the existence of dark solitons within the quantum droplet. When the width of the quantum droplet is much larger than the width of the dark soliton, the density of the quantum droplet can be considered as the background density for the existence of dark solitons, thus forming a coexisting state of quantum droplets and dark solitons.

## II. Framework

We consider a dipolar Bose gas in quasi-1D configuration, with spatial confinement along the longitudinal direction $x$ being finite with a length of $L$, and we investigate $N$ dipolar bosons constrained in the transverse directions $y$ and $z$ within a strongly confining harmonic trap of frequency $\omega_\perp$. These bosons are all polarized along the $x$-direction. And we can introduce an aspect ratio $\sigma = l_\perp/L$ to further characterize the system, where $l_\perp = \sqrt{\hbar/m\omega_\perp}$.

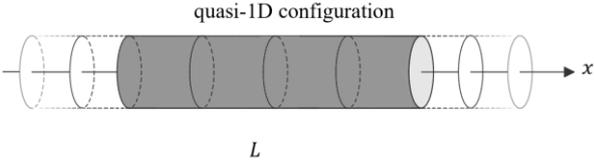

Fig. 1. Quasi-1D geometric model.

A dipolar BEC is subject to two interparticle interactions, the short-range repulsive contact interaction and the long-range attractive dipole interaction. The quasi one-dimensional effective potential is given by [14, 32]

$$V_{\text{eff}}(x) = V_{\text{sr}}(x) + V_{\text{dd}}(x), \quad (1)$$

the quasi-1D dipole potential is

$$V_{\text{dd}}(x) = -\frac{\mu_0 \mu_D^2}{2\pi l_\perp^2} \frac{v_{\text{dd}}\left(\frac{x}{l_\perp}\right)}{l_\perp} := -g_{\text{dd}} \frac{v_{\text{dd}}\left(\frac{x}{l_\perp}\right)}{l_\perp}, \quad (2)$$

$$v_{\text{dd}}(u) = -\frac{1}{2}|u|$$
$$+ \frac{\sqrt{2\pi}}{4}\left(\frac{|u|}{\sqrt{2}}\right) e^{\frac{u^2}{2}} \text{Erfc}\left(\frac{|u|}{\sqrt{2}}\right), \quad (3)$$

where $\mu_0$ and $\mu_D$ denote the permeability of vacuum and the atom magnetic moment, respectively. The short-range contact potential can be written as

$$V_{\text{sr}} = \frac{\hbar^2 a}{m l_\perp^2} \delta(x) := g\delta(x), \quad (4)$$

where the scattering length $a > 0$ can be tuned using Feshbach resonances. We set the units of length, momentum, and energy as $L$, $\hbar/L$ and $\hbar^2/mL^2$ respectively. To further simplify the form of $V_{\text{dd}}(x)$, we define $\sigma = l_\perp/L$ and $v_{\text{dd}}^\sigma(x) := (1/\sigma)v_{\text{dd}}(x/\sigma)$, allowing us to rewrite the effective potential.

$$V_{\text{eff}}(x) = g\delta(x) - g_{\text{dd}} v_{\text{dd}}^\sigma(x). \quad (5)$$

Setting the ratio of the coupling coefficients of the two interactions as $f_{\text{dd}} = g_{\text{dd}}/g_{\text{sr}}$, allows for the analysis of the competition between the two interactions on the system state by considering different values of $f_{\text{dd}}$, especially when $f_{\text{dd}}$ varies around 1.

The Lieb-Liniger (LL) model was proposed to solve one-dimensional Bose gases [10], that considered a one-dimensional Bose gas interacting through a repulsive $\delta$ function potential, providing the general solution form of the system and the relationship between the ground state energy $E_0$ and the nontrivial parameter $\gamma$.

$$E_0 = N\rho^2 e(\gamma), \quad (6)$$

where, $N$ represents the number of particles, $e$ is a monotonically increasing function of $\gamma$, where $\gamma = c/\rho$ with $\rho$ being the density and $2c$ representing the strength of the delta function. When $\gamma = \infty$, due to the impenetrability of particles, the Girardeau result can be derived. When $\gamma = 0$, the non-interacting Bose gas can be derived, while for small $\gamma$, Bogoliubov's perturbation theory has been shown to be very effective.

We can defines the pressure [33]

$$P_{\text{LL}}\left(\frac{N}{L}\right) = -\frac{\partial E_0\left(\frac{N}{L}\right)}{\partial L}$$

$$= \frac{\hbar^2}{2m}\frac{N^3}{L^3}(2e_{\text{LL}}(\gamma) - \gamma \bar{e}_{\text{LL}}(\gamma)), \quad (7)$$

the energy required to add a particle to the system is denoted as the chemical potential

$$\mu = \frac{\partial E_0}{\partial N} = \rho^2\left(3e_{\text{LL}} - \gamma\frac{de_{\text{LL}}}{d\gamma}\right), \quad (8)$$

and the potential energy of each particle is

$$v = \frac{c}{N}\frac{\partial}{\partial c}E_0 = \rho^2\gamma\frac{de_{\text{LL}}}{d\gamma}, \quad (9)$$

the kinetic energy of each particle is represented as

$$\tau = \frac{1}{N}E_0 - v = \rho^2\left(e_{\text{LL}} - \gamma\frac{de_{\text{LL}}}{d\gamma}\right). \quad (10)$$

For different (large and small) values of $\gamma$, the asymptotic forms of the function $e_{\text{LL}}$, the chemical potential $\mu_{\text{LL}}$, the potential per particle $v$, and the kinetic energy $\tau$ also differ slightly. For a large $\gamma$, there is

$$e_{\text{LL}} = \frac{1}{3}\pi^2\left(\frac{\gamma}{\gamma+2}\right)^2, \quad \mu_{\text{LL}} = \frac{3\gamma+2}{\gamma+2}\rho^2 e_{\text{LL}},$$

$$v = \frac{4}{\gamma+2}\rho^2 e_{\text{LL}}, \quad \tau = \frac{\gamma-2}{\gamma+2}\rho^2 e_{\text{LL}}. \quad (11)$$

The monotonically increasing function $e_{\text{LL}}$ is approximated [14] as

$$e_{\text{LL}} = \frac{gN(N-1)}{2}\frac{|\psi|^6}{|\psi|^2 + \frac{3g}{N\pi^2}}, \quad (12)$$

thus, the approximate energy is obtained:

$$E_{\text{LL}} = \int\left[\frac{N}{2}|\nabla\psi|^2 + \frac{gN(N-1)}{2}\frac{|\psi|^6}{|\psi|^2 + \frac{3g}{N\pi^2}}\right]dx$$

$$-\frac{g_{\text{dd}}N(N-1)}{2}$$

$$\int |\psi(x)|^2 v_{\text{dd}}^\sigma(x - \bar{x})|\psi(\bar{x})|^2 \, dx d\bar{x}, \quad (13)$$

where $\int dx\, |\psi(x)|^2 = 1$. Then, we arrive at a new equation by using variational:

$$\mu\psi(x) = -\frac{N}{2}\frac{\partial^2\psi(x)}{\partial x^2} + f_{\text{LL}}[\psi(x)]$$

$$- g_{\text{dd}}N(N-1)$$

$$\int d\bar{x}\, v_{\text{dd}}^\sigma(x - \bar{x})|\psi(\bar{x})|^2\psi(x), \quad (14)$$

where $f_{\text{LL}}[\psi(x)] = \delta e_{\text{LL}}/\delta\psi^*$, and $\mu$ is a Lagrange multiplayer. It is worth noting that when $g \to 0$, the above equation can revert to the standard GPE; and when $g \to \infty$ (Tonks-Girardeau limit), the above equation can be derived to the equation in Ref. [10, 34].

## III. The transition of dark soliton and quantum droplet

When the appropriate $f_{\text{dd}}$ is selected and kept constant, transitioning from bright solitons to quantum droplets can occur by only adjusting the atom number $N$. Fig. 2 shows the evolution from bright soliton states to quantum droplet states as the atom number increases for $f_{\text{dd}} = 20$. Bright soliton states have a maximum density, which is equal to the density of stable quantum droplets. Once the maximum density is reached, further increasing the number of atoms will only increase the width of the quantum droplets, without changing the density.

When a larger atom number $N$ is chosen as a constant value while adjusting the parameter $f_{\text{dd}}$, the system can still exhibit bright soliton states and quantum droplet states. However, unlike the case where the atom number is adjusted for a selected $f_{dd}$ value, in this case, there is no density crossover between the bright soliton states and quantum droplet states. When $f_{dd} > 1$, the system exhibits a net attraction, with the ground state having negative energy, leading to the formation of self-bound states akin to bright soliton states as described in [14]. On the other hand, when $f_{dd} < 1$, the system shows a net repulsion, yet atoms can still form self-bound states. In the extreme case of $f_{dd} = 0$ ($g_{dd} > 0$, $g_{sr} \to \infty$), self-bound states can still be formed through modifications [15]. Similar to self-bound droplet states and self-bound bright soliton states in dipolar systems and Bose-Bose mixtures [35, 36], Ref. [14] identifies the values of $f_{dd}$ for quantum droplets and bright solitons in quasi-1D dipolar Bose gases as $f_{dd}^{\{QD\}} = 0.9$ and $f_{dd}^{\{BS\}} = 20$.

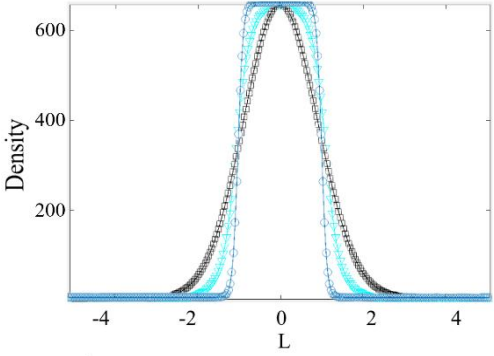

Fig. 2. Bright soliton to quantum droplet state. With an increase in the number of particles, the density of self-bound bright soliton states gradually increases. As the number of particles stabilizing the self-bound droplet state increases, the droplet density remains constant and the particle distribution becomes more uniform.

## IV. The coexistence of dark soliton and quantum droplet

The dimensionless LLGPE under the strong contact limit $\gamma \to \infty$ and zero-range dipole interaction limit is given by [34]

$$i\hbar\dot{\psi} = \frac{\hbar^2}{2m}\psi'' + \pi^2|\psi|^4\psi - g_{dd}|\psi|^2\psi, \quad (15)$$

where $\psi = \psi(x,t)$ denotes the wave function. Upon immediate observation, the segment of this equation governing short-range interactions mirrors that of the Kolomeisky equation [33]. Conversely, the dipolar component adopts the nonlinear form characteristic of the GPE. This approximation is effective when the interaction range σ is smaller than the typical length scale for density variations but significantly larger than the average interparticle distance. As stated in Ref. [37], the solitonic solution in the repulsive dipolar gas under this condition converges to that in a gas with only contact interactions.

By plug the function $\psi_{ds}(x,t) \coloneqq \sqrt{\rho(\xi)}e^{i\phi(\xi)}e^{-i\mu t/\hbar}$, where $\xi = x - vt$, $v$ is the soliton velocity. Then, a complex equation can be obtained [15]

$$\mu\sqrt{\rho} - i\hbar v(\sqrt{\rho})' + \hbar v \psi' \sqrt{\rho}$$
$$= -\frac{\hbar^2}{2m}(\sqrt{\rho})'' - i\frac{\hbar^2}{m}\psi'(\sqrt{\rho})'$$
$$- i\frac{\hbar^2}{2m}\psi''\sqrt{\rho} + \frac{\hbar^2}{2m}(\psi')^2\sqrt{\rho}$$
$$+ \frac{\hbar^2\pi^2}{2m}\sqrt{\rho}^5 - g_{dd}\sqrt{\rho}^3, \quad (16)$$

which can be split into the real and imaginary parts. After separating, dark soliton solutions can be obtained:

$$\rho(\xi) = \rho_\infty - \frac{(\rho_\infty - \rho_{\min})(1+D)}{1 + D\cosh(W\xi)}, \quad (18)$$

$$\psi(\xi) = \frac{2mv(D+1)\left(\frac{\rho_{\min}}{\rho_\infty} - 1\right)}{\hbar DW\sqrt{1-a^2}}$$
$$\arctan\left(\frac{(a-1)\tanh\left(\frac{W\xi}{2}\right)}{\sqrt{1-a^2}}\right), \quad (19)$$

where $a := \frac{\rho_{min}}{\rho_\infty} + \frac{\rho_{min}}{D\rho_\infty} - 1$, $D := \frac{\rho_{min}-\rho_1}{2\rho_\infty - \rho_1 - \rho_{min}}$,

$W := 2\sqrt{\frac{\pi^2}{3}(\rho_\infty - \rho_{min})(\rho_\infty - \rho_1)}$.

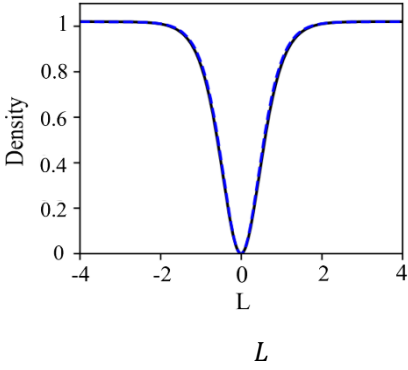

Fig. 3. Dark soliton density profiles for.

## IV. Conclusion

In summary, the results presented in this work confirm the presence of dark solitons in dipolar Bose gases with strong contact interactions. The study focuses on the regime of strong contact interactions where quantum droplets coexist with dark solitons in quasi-1D. By considering the nonlocal LLGPE model, a competition between quintic and cubic nonlinearities is observed. In the case of infinite contact interaction strength and zero-range dipolar interactions, an analytical solution for dark solitons is derived. The width of motionless solitons diverges at a certain threshold, leading to ultrawide solitons that are experimentally observable in large quasi-1D systems. The research on solitons and quantum droplets still holds great potential, and future studies will continue to expand this theory.